\documentclass[fleqn,twocolumn]{openjournal}
\usepackage[T1]{fontenc}
\usepackage{ae,aecompl}
\usepackage{graphicx}   

\usepackage{natbib}
\usepackage{ulem}
\usepackage{amsmath}    
\usepackage{amssymb}    
\usepackage{natbib}
\usepackage{color}      
\definecolor{dgreen}{rgb}{0,0.5,0}
\definecolor{dp}{rgb}{0.5,0,0.5}
\definecolor{magen}{rgb}{0.79,0.08,0.48}
\definecolor{darkred}{rgb}{0.65,0.06,0.37}
\usepackage{booktabs}
\usepackage{multirow}
\usepackage{makecell}

\newcommand{\wfi}{WFI2033}
\newcommand{\iptf}{iPTF16geu}
\newcommand{\hc}{H0LiCOW}
\newcommand{\pix}{\textsc{PixeLens}}
\newcommand{\glass}{\textsc{glass}}
\newcommand{\simgt}{\hbox{\,\rlap{\raise 0.425ex\hbox{$>$}}\lower 0.65ex\hbox{$\sim$}\,}}
\newcommand{\simlt}{\hbox{\,\rlap{\raise 0.425ex\hbox{$<$}}\lower 0.65ex\hbox{$\sim$}\,}}

\shorttitle{Mass models of {WFI 2033-4723}}
\shortauthors{Barrera et al.}

\begin{document}
\nocite{*}

\title[Sample title]{Bridging the gap between simply parametrized and free-form pixelated models of galaxy lenses: the case of WFI 2033-4723 quad}

\author{Bernardo Barrera}
\affil{ Escuela de Ingeniería y Ciencias, Tecnológico de Monterrey, Ave. Eugenio Garza Sada 2501, Monterrey, N.L. 64849, México}
\email{bern.barr@gmail.com}
\author{Liliya L.R. Williams}
\affil{School of Physics and Astronomy, University of Minnesota, 116 Church St SE, Minneapolis MN 55455, USA}
\email{llrw@umn.edu}
\author{Jonathan P. Coles}
\affil{Physik-Department T38, Technische Universit\"at M\"unchen, Ernst-Otto-Fischer-Stra\ss e 8, 85748 Garching, Germany}
\and
\author{Philipp Denzel}
\affil{Institute for Computational Science, University of Zurich, CH-8057 Zurich, Switzerland\\
Physik-Institut, University of Zurich, CH-8057 Zurich, Switzerland}


\begin{abstract}
We study the radial and azimuthal mass distribution of the lensing galaxy in WFI 2033-4723. Mindful of the fact that modeling results depend on modeling assumptions, we examine two very different recent models: simply parametrized (SP) models from the \hc~collaboration, and pixelated free-form (FF) \glass~models. In addition, we fit our own models which are a compromise between the astrophysical grounding of SP, and the flexibility of FF approaches. Our models consist of two offset parametric mass components, and generate many solutions, all fitting the quasar point image data.  Among other results, we show that to reproduce point image properties the lensing mass must be lopsided, but the origin of this asymmetry can reside in the main lens plane or along the line of sight. We also show that there is a degeneracy between the slope of the density profile and the magnitude of external shear, and that the models from various modeling approaches are connected not by the mass sheet degeneracy, but by a more generalized transformation. Finally, we discuss interpretation degeneracy which afflicts all mass modeling: inability to correctly assign mass to the main lensing galaxy vs. nearby galaxies or line of sight structures. While this may not be a problem for the determination of $H_0$, interpretation degeneracy may become a major issue for the detailed study of galaxy structure. 
\end{abstract}

\maketitle

\section{Introduction}\label{intro}

Gravitational lensing is the best way to obtain the detailed projected mass distribution of massive galaxies at intermediate redshifts, acting as lenses for background sources. Two main types of lens mass models are used in the literature, simply parametrized (SP) and free-form (FF).   

The motivation for using SP models is that the central---few percent of the virial radius---regions of galaxies are mostly in equilibrium, as evidenced by observations and simulations, and so can be well approximated by power law density profiles, or other simple parametric functions. The influence of outside mass is parametrized by external shear, added to the lensing potential. (Shear is similar to tides and has no associated mass.) The motivation for using FF modeling approach is that SP models are likely too idealized to represent real galaxies, and may not capture the full range of mass features influencing image properties. FF methods do not use simple parametric profiles and instead allow lensed image properties considerable freedom in determining the mass distribution of the lensing galaxy. The drawback is that these can allow unphysical models to be generated and become part of the ensemble average. \pix~\citep{sah04} and \glass~\citep{col14,col18} pixellate the lens plane and let each mass pixel vary independently, with a few weak superimposed constraints to discourage unphysical mass models.

Most published works to date assume SP forms as the galaxy lensing model. The most common models are elliptical power law density profiles with slopes near `isothermal', or two co-centered mass components, for example, Sersic \citep{ser68} to represent baryons, and NFW \citep{nav97} to represent dark matter. 

Some years ago it became clear that SP mass models cannot account for the observed image flux ratios in many multiply imaged quasar systems \citep{fal97}. At optical wavelengths, image fluxes are most likely affected by microlensing by stars in the main lensing galaxy. Since microlensing operates on much smaller spatial and mass scales, it can be studied separately, and is usually disregarded when the focus is on the galaxy-scale macromodel. Flux ratios at longer wavelengths are not subject to microlensing, but can be affected by dark matter substructure \citep{mao98,met01,chi02}, or other mass features in the macromodel. To explore the latter, attempts were made to use smoothly perturbed macromodels \citep{tro00,eva03}. While in some quads these models could account for flux anomalies, they often resulted in unphysical projected iso-density contour shapes \citep{big04}. {The substructure predicted by the Cold Dark Matter Universe model with the Cosmological Constant ($\Lambda$CDM)}, or line of sight (LoS) structure \citep{gil19,nie20}, and, in some cases, edge-on disks \citep{hsu16} are probably responsible for some, but not all flux anomalies, and there remains a tension between predictions and observations \citep{xu15,hsu18}. 

Given the shortfalls of SP and FF modeling strategies, a different approach may be warranted. {(Even though no approach is perfect!)} In this paper we use macromodels which are, in a sense, a compromise between SP and FF strategies. Similar to SP models, our models consist of simply parametrized mass components, each with its own independently varying normalization, core, density slope, ellipticity and position angle. Unlike the SP models, the centers of our two components are allowed {to be offset}. A superposition of such components usually results in lopsided mass distributions, but the center of mass of our models is required to coincide with the center of light to within astrometric precision, which is not the case for SP models.

Because the number of our free parameters exceeds the number of constraints, our method, like FF, produces a large number of solutions, instead of just one macromodel with statistical uncertainties, as is the case with standard SP.  The reason for multiple solutions is the existence of lensing degeneracies, i.e., there is no guarantee of uniqueness. In fact, as was pointed out by \cite{wag19}, lensing observables provide us with local lens properties only, the rest we obtain by filling in the information gaps with our assumptions, or non-lensing observables.  So to obtain robust model uncertainties, one needs a wider range of models, which implies a larger number of adjustable parameters. 

Thus, we borrow from both the SP and FF approaches. From SP we take the simple shape of the building blocks of galaxy structure, which are physically motivated. From FF we take the idea of multitude of solutions, all reproducing image observables within observational error, and therefore connected by lensing degeneracies.

The motivation for the two offset component model comes from recent work. Based on analysis of a population of galaxy quads, \cite{gom18} conclude that some fraction of lens galaxies needs to be lopsided. \cite{nig19} perform modeling of a handful of individual quad lenses, and conclude that off-centered components are required. \cite{williams2020two} find that lopsided mass maps can bring quadruply imaged Supernova Type Ia \iptf~into agreement with microlensing constraints, while still allowing the galaxy profile to be steep at large radii, and have coincident light and mass centroids. 

In this paper we model WFI 2033-4723 ({hereafter referred to as WFI2033}), a quad lens {first reported by \cite{morgan2004wfi}}. Our motivation for choosing it is that it is one of six \hc~systems, and its quasar point images and lens light center are poorly fit by power law and composite models \citep{rusu2020h0licow}.

The recent galaxy-lens modeling literature is heavily dominated by the goal of measuring $H_0$. In this paper we are interested in placing constraints on the galaxy mass distribution. We elaborate on our choice of constraints in  Section~\ref{sec:data}, and in Section~\ref{sec:models} we describe our fitting procedure, {which generates models that fit quasar image data to observational precision. The paper's goal is to determine what can currently be said about the structure of the lens galaxy in \wfi. Two different aspects of this goal are addressed in Sections~\ref{sec:constraints} and \ref{sec:comparison}. In Section~\ref{sec:constraints} we present 3 sets of models, built with 3 different sets of constraints, which allows us to evaluate their respective model constraining power. In Section~\ref{sec:comparison} we compare our models with published models, identifying similarities and differences. Section~\ref{sec:conc} summarises our findings.}

\section{WFI 2033-4723 and data constraints}\label{sec:data}

{\wfi} is a quadruply imaged quasar with source and lens redshifts of $z_{\textrm{src}} = 1.662$ and $z_{\textrm{lens}} = 0.6575$ \citep{sluse2019h0licow}.  We use image-positional data from COSMOGRAIL X \citep{sluse2012cosmograil}, and time delay data from COSMOGRAIL XVIII 
\citep{bonvin2019cosmograil}. While most current papers on galaxy scale lenses focus on $H_0$, our interest is in the lens plane mass distribution. Because we do not want to bias our results with a fixed value of $H_0$, we do not use the actual time delays, but only their ratios, which eliminates the dependence on $H_0$. 
Thus, we define the two independent ratios $r_{1} = \frac{\Delta t_{B-A2}}{\Delta t_{B-A1}}=1.0304\pm 0.1329$ and  $r_{2} = \frac{\Delta t_{B-C}}{\Delta t_{B-A1}}=1.6409\pm 0.1243$, which are used to fit our model parameters to time delay data.  The dependence on other cosmological parameters, mostly $\Omega$'s, remains, but it is much weaker.

One of the sets of models we present uses three image flux ratios, obtained from narrow line region of the source quasar, which is unaffected by stellar microlensing \citep{nie20,gil20}, and hence probes either $\Lambda$CDM substructure, or macro-scale potential of the galaxy, or a combination of the two. The three flux ratios are, 
$f_1=F_B/F_{A1}=0.50\pm 0.050$, 
$f_2=F_{A2}/F_{A1}=0.65\pm 0.055$, and 
$f_3=F_C/F_{A1}=0.53\pm 0.048$.

In addition to constraints from the quasar point images, many papers use the lensed image of the extended galaxy hosting the source quasar, i.e., the extended ring. Also, to constrain the very central region, well within the image ring, analyses often use velocity dispersion of the lensing galaxy. In the following subsections we argue that not all data is of equal value for modeling, since some have intrinsically lower quality observations, and others are more reliant on prior assumptions. 

\subsection{Constraints from point images}\label{sec:pointims}

Quasar image positions require no {additional assumptions} or  modeling priors: point images arise from the same very small, practically point-like location in the source plane. With HST observations, these have astrometric accuracy of $\sim 0.003"$. In the absence of microlensing, quasar image time delay also require no priors. Fortunately, the effect of microlensing is small, similar or smaller in magnitude to the time delay measurement uncertainties \citep{rusu2020h0licow}. So the use of time delays in mass modeling of \wfi~does not require priors. 

\begin{table}
\centering
\caption{Quasar image data used in the 3 levels of modeling}\label{tab:groups}
\begin{tabular}{c|ccc}
   \hline
   Group &  positions & time delay ratios & flux ratios \\
   \hline
     A   & \checkmark        & \sffamily X            & \sffamily X \\     
     B   & \checkmark        & \checkmark            & \sffamily X \\
     C   & \checkmark        & \checkmark            & \checkmark \\
   \hline
\end{tabular}
\end{table}

Physical quasar sizes in the mid-IR to radio wavelengths, or in the narrow line regions are too big to be influenced by stellar microlensing. Instead, they are affected by anything from $\Lambda$CDM substructure and LoS halos, to macro-scale mass features in the lens galaxy. From the data alone it is not possible to differentiate between these possibilities. The method adopted in the literature is to assume a form for the lensing galaxy, either analytical or extracted from simulations, and ascribe deviations from model predictions to small-scale substructure. The reliability of this approach depends on how well the model describes the real lens on length and mass scales that are a fraction of the Einstein radius and typical $\kappa$ values of the main lens.

In this paper we do three levels of mass reconstruction, {each incorporating successively more constraints}. Our Group A models use image positions only, {Group B models use both image positions and time delay ratios}, 
and finally Group C further adds image flux ratios to {image positions and time delay ratios}; see Table~\ref{tab:groups}.
In other words, Groups A and B reconstructions assume that flux ratios are due to small-scale substructure and therefore cannot be used to constrain the macro models. Group C reconstructions assume that flux ratios are the result of perturbations in the macro-model, and so are included as a constraint. 
 
\subsection{Constraints from extended rings}

The galaxies hosting the source quasars are lensed into extended rings that cover hundreds or even thousands of HST CCD pixels, and sample the lensing potential over a considerable portion of the lens plane. However: (i) each pixel is 0.03"-0.05", which is worse than the astrometric precision of point images by a factor of 10 in linear scale, or a factor of 100 in area, and (ii) their surface brightness is usually low, and has to be separated from the light of the lensing galaxy, leading to low signal-to-noise {ratio}.  Furthermore, (iii) a lensing system comprises a lens and a source, and the two are somewhat interdependent in modeling, leading to lens-source degeneracy. Since we do not know the true appearance of the extended source (unlike in the case of a point source), complexity in the structure of the lens can to some degree be attributed to the source, or vise versa. 

Usefulness of rings in modeling has been debated in the literature. {\cite{koch01rings} was one of the first papers to analyse the information contained in the extended rings, and argue that these could be very useful, even ``revolutionary" in constraining lens mass models and breaking degeneracies. However, their modeling made simplifying assumptions about the shape of the lens and the source. In particular, the isophotes of the central region of the source were taken to be ellipsoidal, while the actual lens-reconstructed sources have non-negligible deviations from ellipticity \citep[see images of reconstructed sources in ][]{veg10a,veg10b,veg12}. }
\cite{sah01} showed that the 4 point images of a quad lens already determine the overall structure of the extended host galaxy image. Many other papers, for example, the \hc~collaboration \citep{suy17} argue that the details of the ring structure help constrain the fit. \cite{wal18}, using a somewhat optimistic set up to represent the ring in synthetic lenses---an array of point images---conclude that the ring provides useful constraints in about half of the cases. \cite{gom21}, also using an array of point images in 10 different synthetic lens types, showed that including rings in the estimation of $H_0$ improved it for 2 out of 9 simpler lens types, but did not improve it for the set of lenses with most realistic mass distributions. The utility of rings can vary between systems, or even analyses. \cite{mor20} found that in \iptf~the extended ring did not help in modeling, while \cite{williams2020two} found that some mass models did differ in their predicted rings, if sufficient resolution and high S/N were available. {\cite{denzel2021hubble} used a different method to incorporate extended ring information.  After generating 1000 individual \glass~models, they selected a subset that reproduced the extended rings reasonably well. However, the predicted $H_0$ distribution from this subset did not differ from that of the whole sample.}


\subsection{Constraints from lens galaxy kinematics}

Kinematics of the lens galaxy, namely the central velocity dispersion has been used extensively in \hc~papers. However, the use of velocity dispersion relies on assumptions. The commonly used Jeans analysis assumes geometrical sphericity, with a velocity anisotropy profile form parametrized by a scale length. The effect of the assumptions on kinematics can be gauged from the results of \cite{bir16} in the case of the quad RXJ 1131-1231. They find that the choice of priors on velocity anisotropy impacts their cosmological inference; they derive $H_0=86.6^{+6.8}_{-6.9}$~km$\,$s$^{-1}$ vs. $H_0=74.5^{+8.0}_{-7.8}$~km$\,$s$^{-1}$  
depending on the prior.  While this does not say what it would imply for the study of the lens galaxy mass distribution, it is reasonable to assume that the results will be affected. \cite{gom20} study the effect of including kinematics in their synthetic two-offset-component mass models, which are then fitted with single power laws to emulate common practice. They find that Jeans modeling pulls the best fit lens models in the direction of steeper density profiles than those of the true lens. They attribute this bias to the fact that the fitted models can never be an exact match to the projected mass distribution of the lens. {Even though this study focused on $H_0$,
its results suggest that the inference on galaxy structure depends on the model used for kinematics.}

\begin{table*}
\centering
\caption{Accuracy in the recovery of $H_0$ in Rungs 1 and 2 of the TDLMC for teams that used SP methods. }\label{table1}
\begin{tabular}{cccccccc}
    \toprule
    \multirow{2}{*}[-4pt]{{Good Team}} &
    \multirow{2}{*}[-4pt]{{extended ring}} &
    \multirow{2}{*}[-4pt]{{kinematics}} &
    \multicolumn{2}{c}{{Rung 1}} &
    \multicolumn{2}{c}{{Rung 2}} & 
    \multicolumn{1}{c}{{Rung 3}} \\
    \cmidrule(lr){4-5}
    \cmidrule(lr){6-7}
    \cmidrule(lr){8-8}
    &&& accuracy & tally & accuracy & tally & tally \\
    \midrule
 Student-T   &  yes   &  no   &{\bf 0.3 -- 2.5\%}& 2/6  &3.7 -- 8\%        & 0/5 & 0/12\\
 EPFL        &  yes   &  yes  &     7.5\%        & 0/1  &{\bf 1.74 -- 2\%} & 2/2 & 0/6 \\
 Rathnakumar &  no    &  yes  &   3.9 -- 6.3\%   & 0/3  &{\bf 1.5} -- 4.8\% & 0/3 & -- \\    
    \bottomrule
\end{tabular}
\end{table*}

\subsection{Constraints from extended rings and kinematics in simple lenses}

The studies discussed above use either real lens systems, or synthetic ones with two mass components. The cleanest test of the usefulness of extended rings and kinematics is to use very simple mock lens mass distributions, and fit them with exactly the same type of mass models. That way, one is eliminating the possibility that unknown complexities in the lens mass distribution are reducing the usefulness of the information contained in rings and kinematics. 

The recent Time Delay Lens Modelling Challenge \citep[TDLMC,][]{ding2020time} is very useful for that purpose. The goal of TDLMC was to estimate $H_0$, not to study the galaxy structure, but their conclusions are still relevant. The analysis was completely blinded.

Of particular interest are Rung 1 and 2 of TDLMC, where the challenge lenses, created by the ``Evil" Team, and unknown to the modelers, or ``Good" Teams, were elliptical power laws with density slope within $0.1$ of isothermal, i.e., $\rho_{3D}\propto r^{-2\pm 0.1}$, magnitude of the external shear was $\gamma\leq 0.05$, and external constant $\kappa\leq |0.025|$. 

As it turned out, three of the Good Teams used the same parametric assumptions to model the lenses as the Evil Team's design of the true ones: elliptical power laws with external shear. Furthermore, the method of calculating the velocity dispersion using the spherical Jeans equation, and the parametrization of the velocity anisotropy were exactly the same for the Evil Team as well the Good Teams that used kinematics. 

The Evil Team set the goals for the recovered accuracy and precision of $H_0$  at 2\% and 6\%, respectively. The results are summarized in Table~\ref{table1}, where the first three columns give the names of the Good Teams and the data they used in their modeling, in addition to that from the quasar point images. The table shows the magnitude of the accuracy for Rung 1 and 2 reconstructions, and highlights in bold values that were within the stated goal of the Challenge. Columns labeled ``tally" give $n/m$, where $m$ is the total number of reconstructions submitted by that Good Team, and $n$ is the number of those that satisfied the accuracy and precision goal. 

Overall, Team EPFL, which used both extended ring and lens velocity dispersion information performed the best, but they failed to meet the goal of the simplest Rung 1. The other two Good Teams, one not using kinematics, the other not using extended ring, did not do much worse.  (In Rung 3, which used lenses extracted from Illustris simulation, and suffered from some issues, neither EPFL not Student-T met the goal of the Challenge.)


Based on Rungs 1 and 2 we conclude that even if the true and the fitted lens density profiles have the same simple functional form, knowledge of the lens velocity dispersion and/or the extended ring does not always guarantee accuracy or precision, or give a team a definitive advantage.  In the real universe, the true mass distribution is not known, complicating the situation, and possibly rendering kinematics and rings even less useful. However, in the future, spatially well resolved kinematic data may prove to be valuable.


Finally, in the case of \wfi~there is possible evidence that the assumptions going into the kinematics modeling and/or the extended ring are in tension with the astrometric data from quasar point sources. The best fitting models of \cite{rusu2020h0licow}, who use kinematics and extended ring, have image rms of $0.01"-0.025"$, compared to HST astrometric precision of $\sim 0.003"$. At the same time their modeled lens mass center is offset from the observed light center by $0.02"-0.03"$, while the astrometric precision is ten times better, $0.002"$. 

Therefore we use image positions, time delays, and in separate modeling also include flux ratios that are immune to microlensing. We do not use the extended ring or stellar kinematics.

\section{Lens plane mass models }\label{sec:models}

\subsection{Galaxy model}

As discussed above, the galaxy model presented in \cite{rusu2020h0licow} does not account well for the positions of the four quasar images or the center of the lens galaxy. Our motivation for using a more complex model is to fit the images better, and to do so with models whose mass and light centers coincide.

As argued in the Introduction, there is evidence that lensing galaxies' mass distribution may be lopsided, which leads us to use two superimposed, but off-centered mass components.
Both are individually modelled by \textbf{alphapot} analytical lensing potential \citep{keeton2011gravlens},
\begin{equation}
    \Psi = b\big(s^2 + x^2 + \frac{y^2}{q^2} + K^2xy\big)^{\frac{\alpha}{2}}.
    \label{eq:alphapot}
\end{equation}
The five free parameters in eq.~\eqref{eq:alphapot} capture the shape, size, orientation, and slope of each component: $b$ fixes the mass normalization, $s$ is the core radius size, $q$ and $K$ together define orientation and eccentricity, and $\alpha$ is the potential profile slope. The position angle $\theta$ of each \textbf{alphapot} component is given by
\begin{equation}
    \tan 2\theta = \frac{q^2K^2}{q^2-1},
    \label{eq:positionAngle}
\end{equation}
and the axis ratio, $Q$ is
\begin{equation}
    Q^2=\frac{\cos 2\theta(q^2+1)+(q^2-1)}{\cos 2\theta(q^2+1)-(q^2-1)},
    \label{eq:QQQ}
\end{equation}
which depends on both $q$ and $K$, through eq.~\eqref{eq:positionAngle}. Though the models we use are simply parametrized, the inclusion of non-zero offset between the two mass components makes these considerably more flexible than the standard parametric models, {especially in the azimuthal direction}.

In addition, HST data reveals that the lens is part of a larger galaxy group at $z_{\rm grp}=0.6588$, with velocity dispersion $\sigma = 500 \pm 80 \,\textrm {km s}^{-1}$, and with at least 6 member galaxies within 2 arcseconds of the main lens \citep{wilson2016spectroscopic}. Of the nearby perturbers, other authors have concluded that G2, a bright galaxy at $z = 0.7450$ with $\log M_\ast \sim 11.15$, is the only galaxy with significant impact on the lens modeling \citep{rusu2020h0licow}.
{\ Our reconstructions assume a single lens plane, which contains the main lensing galaxy, as well as G2. Had we placed G2 at its actual, somewhat higher redshift, its effective location on the plane of the sky would be slightly different, and its surface mass density would be increased by about 7\%.}
Galaxy G2 looks round, so we model it as a circularly symmetric \textbf{alphapot} potential with an isothermal density slope, a negligibly small core, and Einstein radius of 1", very similar to the estimate of \cite{rusu2020h0licow}.  These authors allow their modeling procedure to optimize the ellipticity and position angle of G2. In Section~\ref{sec:interp} we will discuss what effect the differences in how to model G2 have on the conclusions.

{The main lensing galaxy in WFI2033 lives in a rather complex local environment: there are two nearby galaxy groups \citep{sluse2019h0licow}, so G2 is not the only nearby perturber. In this paper we account for the combined influence of other galaxies with a single external shear. {The magnitude of the shear is quite large, $\gamma\sim 0.1-0.2$, indicating that the perturbers have a significant influence on the lens.}
A single shear is only an approximation because the actual sky-projected distribution of mass due to nearby galaxies cannot be represented by a single shear axis. Quantifying the limitations of the single shear approach will be an interesting exercise, but is beyond the scope of this paper. We note that the other two models we discuss,  \cite{rusu2020h0licow} and \cite{den2020}, also use single shear to model the lens environment and can be directly compared to ours in that regard. }

\begin{figure*}
    \includegraphics{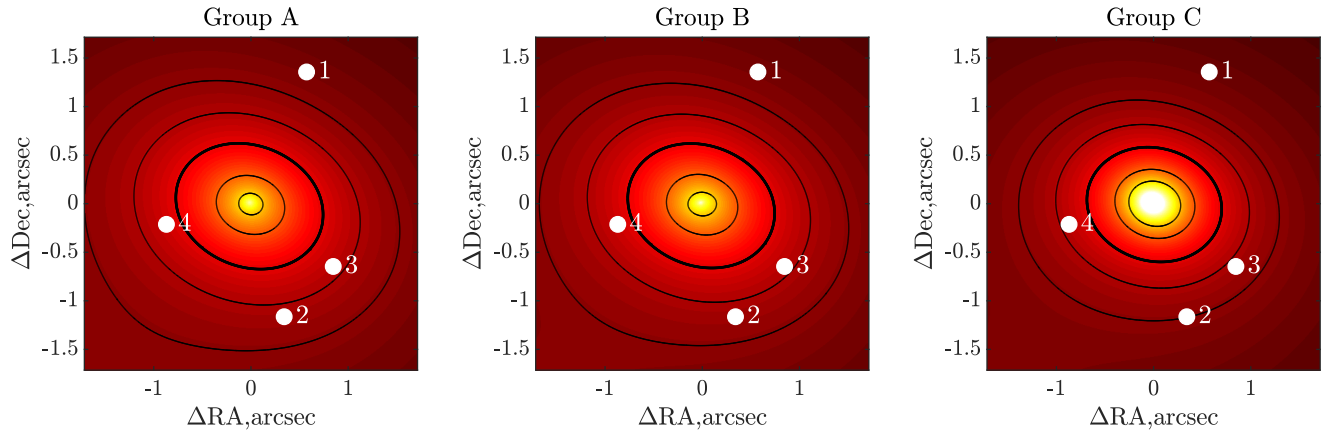}
    \caption{Average convergence $\kappa$ plots for 50 individual models of Group A, B and C (see Table~\ref{tab:groups}), with $\kappa =  0.6, 0.75, 1 ({\rm thick}) , 1.5, 2$ contours shown in black. Images are ordered with respect to arrival time. Galaxy G2 is located at $235^\circ$, measured counterclockwise from the positive $x-$axis. {In standard $\Lambda$CDM, $1"$ at the redshift of the lens corresponds to $6.96$~kpc.}}
    \label{fig:AverageConvergence}
\end{figure*}

\begin{figure}
    \centering
    \vspace{0.2in}
    \includegraphics{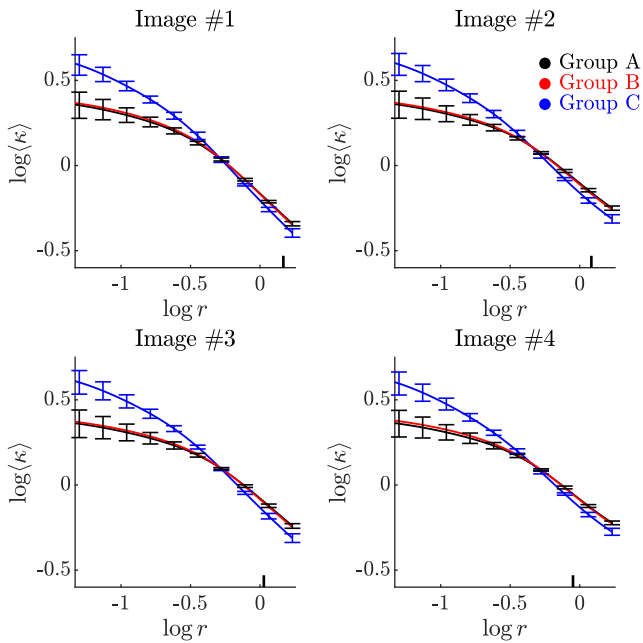}
    \caption{Mass density profiles averaged over 50 models and over $30^\circ$ circle arcs, centered on the lens mass center and opening towards each of the 4 images. Small black vertical line segments along the horizontal axis indicate the locations of the images. Red, black and blue curves represent Groups A (image positions only), B (image positions and 2 time delay ratios), and C (image positions, 2 time delay ratios, and 3 flux ratios) reconstructions respectively. The uncertainties are the rms dispersion of 50 individual models. 
    }
    \label{fig:OurModelsFlux}
\end{figure}

\subsection{Model fitting}

Having established the general form of our lens models, we proceed to find good parametric fits to observational data. 

Each of the two \textbf{alphapot} components described by eq. \eqref{eq:alphapot} is specified by a set of 5 parameters. Thus, a complete mass model consists of 10 free variables and an extra coordinate pair $(x_{\textrm{off}},\,y_{\textrm{off}})$ describing the offset of the secondary with respect to the primary mass component. In addition, each model has an associated external shear magnitude and position angle $\gamma_{ext}$ and $\theta_{\gamma}$, and a coordinate pair for the source position. In total, these constitute 16 free parameters per model. 
To obtain good fits to observational data, we proceed in two steps. First,
the model parameters are chosen randomly from reasonable ranges that restrict the ellipticity parameters from deviating far from circular and delimit the maximum core radius sizes of the compact {(centered on the light center)} and extended {(allowed non-zero offset)} mass components to at around $20\%$ and  $90\%$, respectively, of the smallest image radius. {The limits on the offset coordinates, $(x_{\textrm{off}},\,y_{\textrm{off}})$ were chosen to be small, $\pm 0.02"$, similar to the offset between mass and light centroids in the \cite{rusu2020h0licow} models.} These ranges are shown as dashed boxes in Figure~\ref{fig:Clusters} in Appendix~\ref{app:Clusters}. Although these initialization values are to some extent arbitrary, we echo the conclusion of \cite{williams2020two} in that parameters are able to deviate well beyond these initial regions after further optimization in the second step (see below). Using each of these randomly generated models we then ray trace the observed images back to the source plane; models are kept or discarded depending on the degree to which their backprojected images converge on the source plane. {In this first step we chose to backproject images to the source plane instead of forward lensing sources because the former is computationally cheaper than the latter. Working in the source plane at this stage is sufficient, since we are looking for approximate solutions only. Their lens plane positional rms, defined below by eq.~\ref{eq:chipos}, is not very good, typically $\chi_{\rm pos}\lesssim 50$.} 
It is relatively easy to find many distinct fits. We generate 100 of these models.  The average source plane $(x_s,\,y_s)$ position of the four backprojected images per model, together with the 14 lens model parameters are then used in the second step. 

Next, we use the Downhill Simplex method \citep{nelder1965simplex} to find good solutions, {defined by the merit function,} $\chi^2$, which is calculated as a sum of the following contributions. 
\begin{equation}
    \chi_{\rm pos}^2 = \sum_{i=1}^4 \bigg(\frac{(x_{\textrm{i,m}}-x_{\textrm{i,o}})^2}{\sigma^2_{\textrm{x,i}}} + \frac{(y_{\textrm{i,m}}-y_{\textrm{i,o}})^2}{\sigma^2_{\textrm{y,i}}}\bigg),
    \label{eq:chipos}
\end{equation}
\begin{equation}
    \chi_{\rm cen}^2 = \frac{x_{\textrm{c,m}}^2+y_{\textrm{c,m}}^2}{\sigma^2_{\textrm{c}}},
\end{equation}
\begin{equation}
    \chi_{\rm td}^2 = \sum_{i=1}^2 \frac{(r_{\textrm{i,m}}-r_{\textrm{i,o}})^2}{\sigma^2_{\textrm{r,i}}},
\end{equation}
\begin{equation}
    \chi_{\rm flux}^2 = \sum_{i=1}^3 \frac{(f_{\textrm{i,m}}-f_{\textrm{i,o}})^2}{\sigma^2_{\textrm{f,i}}},
\end{equation}
where subscripts o and m stand for observed and model predicted values.
Image coordinates are $(x_{\textrm{i}},y_{\textrm{i}})$ for images i=1,...~4, time delay ratios are $r_{\textrm{i}}$, for image pairs i=1,~2, flux ratios are $f_{\textrm{i}}$, for image pairs i=1,~2,~3, and $\sigma$'s are their associated uncertainties. The image-positional $\sigma_{\textrm x}=\sigma_{\textrm y}=0.002"$. The values and uncertainties on time delay ratios and flux ratios are given in Section~\ref{sec:data}. The location of the model center of mass is denoted by $(x_\textrm {c,m},\,y_\textrm {c,m})$, with respect to the center of light of the lens, and $\sigma_{\textrm c}=0.002"$ is the separation between the lens galaxy light centers in HST F814W and F160W filters \citep{rusu2020h0licow}. {$\chi^2_{\rm {cen}}$ penalizes the separation between the model center of mass, $(x_\textrm {c,m},\,y_\textrm {c,m})$, and the center of light, which is set at the origin. Note that the coordinate pair $(x_\textrm {c,m},\,y_\textrm {c,m})$ is not the same as the center of the secondary mass component, but is a function of all lens model parameters}.

As described in Section~\ref{sec:pointims} and Table~\ref{tab:groups} we carry out three levels, or Groups of reconstructions. All three of our Groups use $\chi_{\rm pos}^2$ and $\chi_{\rm cen}^2$. In addition, Group B uses $\chi_{\rm td}^2$, and Group C uses all four. These amount to {$n=$10, 12, and 15} independent data constraints, for the three groups, respectively. {For each case, the total $\chi^2$ is given by the sum of the corresponding contributions divided by the number of independent data constraints $n$.}

The Downhill Simplex search works as follows. Each of the 100 models {found in the first step} is thought of as a point in the 16-dimensional space spanned by our parameters. In turn, a pair of points define opposite vertices of a 17-sided simplex in the parameter space. By selecting many random pairings among the 100 models, we construct simplices which shrink and move across the space until a certain degree of convergence is reached around a local minimum. Though not every minimum represents a good solution, this optimization allows us to find many local minima with $\chi\leq 1$. {These correspond to models that reproduce all observables, on average, to within experimental precision. We do not find solutions with negligibly small $\chi$. This is probably a limitation imposed by our parameterization}.  

We obtain 50 models with $\chi\leq 1$ in each of Groups A, B and C, the average convergences for which are shown in Figure~\ref{fig:AverageConvergence}.  

\section{Constraining power of image positions, time delay ratios and flux ratios}\label{sec:constraints}

One of our goals is to determine how additional data from the point images contributes towards constraining the mass models of the lens galaxy.

Figure~\ref{fig:AverageConvergence} shows the mass maps for the 3 Groups of models, summarized in Table~\ref{tab:groups}. The maps include the contribution from galaxy G2, located about $4"$ from the center, at $235^\circ$ measured counterclockwise from the positive $x-$axis. {Including the two time delay ratios as well as} image positions, i.e., comparing Groups A and B, does not appear to change the resulting maps.  Measuring time delays for multiply imaged quasars is a resource-intensive task. In the case of \wfi~it took 14 years on more than one 1-3 meter class telescopes to extract precision of $2-6\%$ \citep{bonvin2019cosmograil}. These results show that at least within the context of our models increased precision may not help in further constraining the mass distribution of this galaxy.\footnote{Note, however, that precision in time delay translates directly to that in $H_0$, so time delays are extremely important for that purpose.}

Another way to compare the mass distributions is presented in Figure~\ref{fig:OurModelsFlux}, which shows density profiles, averaged over $30^\circ$ circle arcs (or wedges, or `pie slices') and over 50 models. These circle arcs have their apex at the galaxy center, and open up in the directions to the 4 images. 
In other words, we are taking the part of the density distribution that lies along the line connecting the lens center and each image.
The Group A (black) and Group B (red) profiles are nearly identical, and well within $1\sigma$ of each other.

The last panel of Figure~\ref{fig:AverageConvergence}, Group C, shows that adding three image flux ratios does significantly impact the solutions. The most notable change is the steeper density profile slope, which is also evident in Figure~\ref{fig:OurModelsFlux}, where the $1\sigma$ uncertainties show that Group C reconstructions are different from those of A and B. {It is not entirely surprising that image magnifications can constrain profile slope. Magnification is image stretching in the radial and tangential directions. Stretching in the tangential direction depends mostly on the image distance from the lens center, while that in the radial direction depends on the profile slope. This result is also consistent with the findings of \cite{ori19}, who use circular power law lenses to show that image flux ratio is critical in determining the lens' density slope.}  

The 4 panels of Figure~\ref{fig:OurModelsFlux} also capture the elongation of the isodensity contours in each of the image directions, and so encode information about the ellipticity of the models. Because there is no significant difference between the black, red, and blue lines in each of the panels, the ellipticity is not affected by the inclusion of flux ratios.

Therefore, at least for the family of models considered here, image flux ratios, if due to the macromodel instead of small $\Lambda$CDM substructure, are significantly more informative than time delay ratios, especially in determining the density profile slope.

\begin{figure*}
    \includegraphics{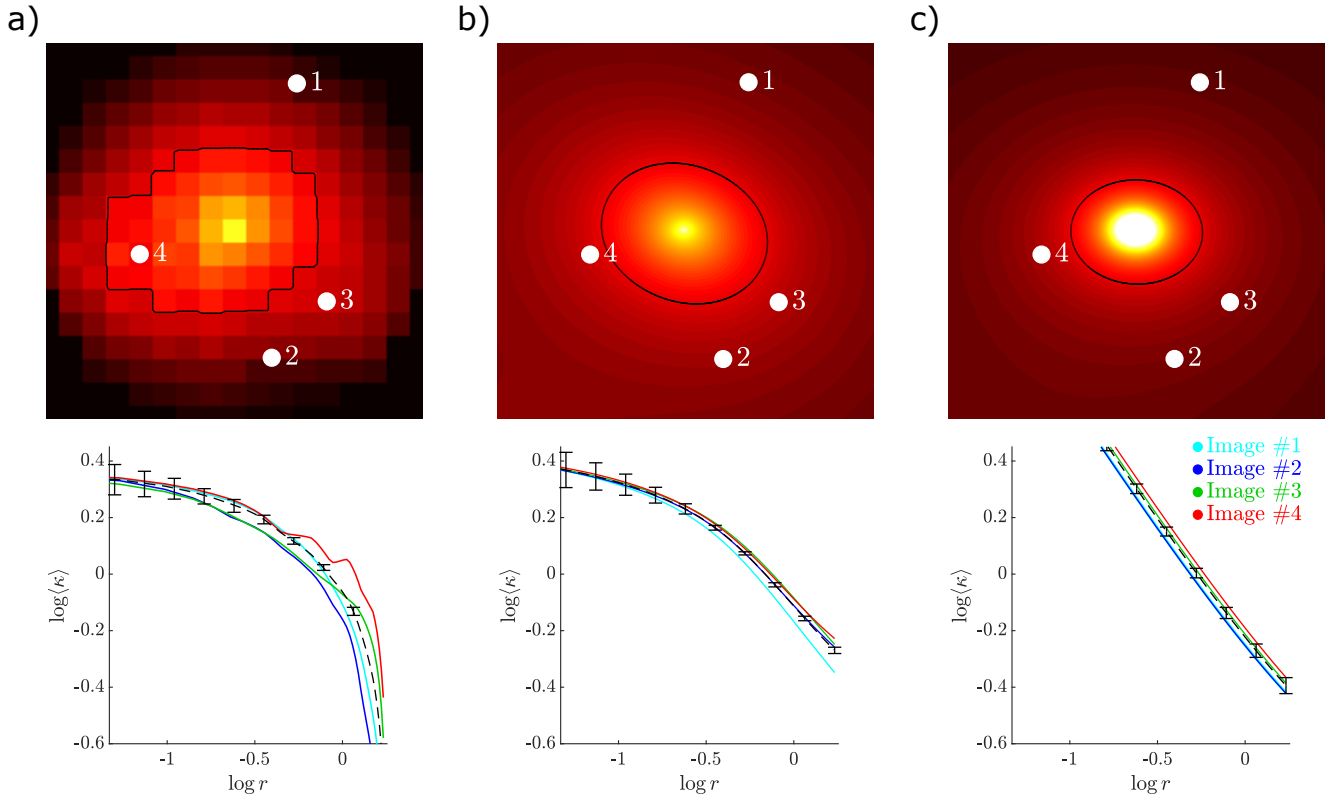}
    \caption{{\it Upper panels:} Average convergence plots for three mass models, with $\kappa=1$ contours shown in black. a) Free-form \glass, b) our Group B models, c) \hc~ SPEMD model. Images are numbered in order of arrival time. Galaxy G2 is located at $235^\circ$, measured counterclockwise from the positive $x-$axis. {\it Lower panels:} The density profiles for the 3 models. For each model we plot profiles in $30^\circ$ circle arcs centered on the lens mass center and opening towards the 4 images. {Our Group B and \glass~profiles are the mean of the logarithm of the density profiles of 50 and 1000 models, respectively, and the error-bars represent the rms scatter between models. For SPEMD the profiles and error-bars are from the \cite{rusu2020h0licow} best fit model from their Table~1.}}
    \label{fig:Comparison}
\end{figure*}

\section{Comparison with Other Published Models}\label{sec:comparison}

We now compare our models to other published results: the free-form \glass~models presented in \cite{denzel2021hubble}, and the {singular power-law elliptical mass distribution (SPEMD)} parametric models presented in \hc~ XII \citep{rusu2020h0licow}.\footnote{We do not consider their composite model because its point image lens plane rms is $2.5\times$ worse than that of SPEMD.} The SPEMD mass distribution is not explicitly plotted in that paper, but we reconstruct it using the parameters in their Table 1. The recovered \glass~and SPEMD mass distributions are shown in Figure~\ref{fig:Comparison}, alongside our own Group B results. {We chose Group B here because it uses the same point image data as the \glass~and SPEMD models.} The $\kappa\!=\!1$ contour is shown in black in all three panels. 

\begin{figure}
    \centering
    \includegraphics{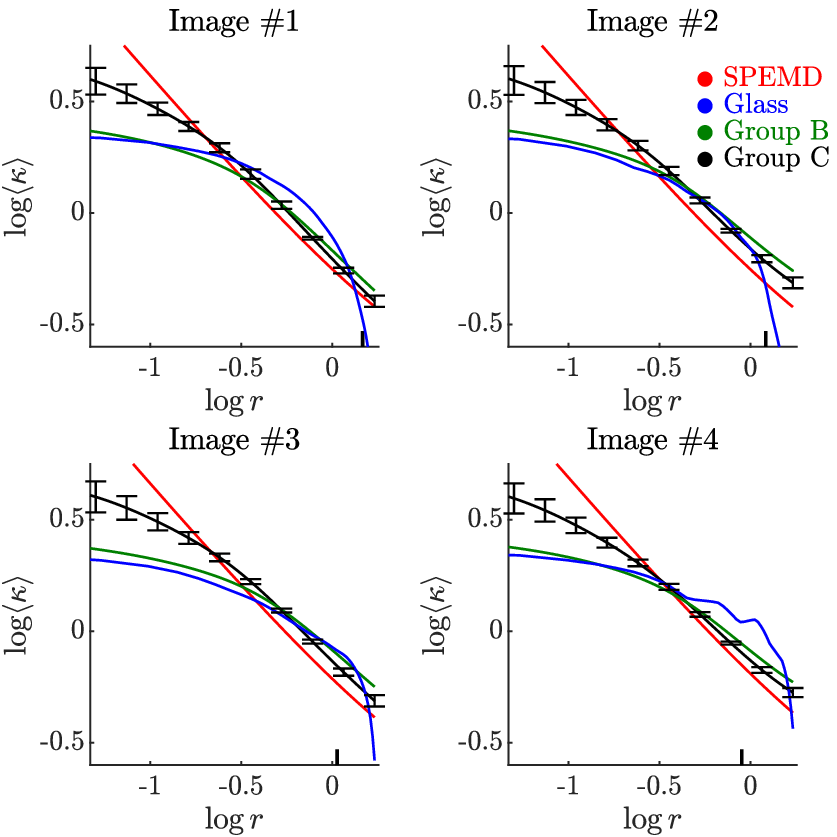}
    \caption{{Density profiles averaged over the corresponding number of models and $30^\circ$ circle arcs, centered on lens center and opening in the directions of the 4 images. Four models are shown in each panel: SPEMD, Glass, and our Group B and C. The number of models is 50 for our Groups B and C, 1000 for \glass, and 1 for SPEMD. (Group A, not shown, is very similar to B.) Small vertical dashes along the $x-$axis mark the locations of the 4 images. To avoid overcrowding only Group C has its error-bar plotted; the error-bars for other models are shown in  Figure~\ref{fig:OurModelsFlux} and the lower panels of Figure~\ref{fig:Comparison}}.}
    \label{fig:mixed}
\end{figure}
\vspace{0.5cm}


\subsection{Radial mass distribution}

\subsubsection{Density profile outside of the very central region}\label{sec:denprof}


Figure~\ref{fig:mixed} shows \glass~(blue), SPEMD (red), and our Groups B (green) and C (black). Group A is not shown here because it is very similar to B. The four panels show density profiles averaged over $30^\circ$ circle arcs and 50 models, with the apex at the lens center and opening towards the four images. Most of these profiles do not look the same. For a given lens system, like \wfi, mass distributions from independent models can look different due to differences in input data {and lensing degeneracies. Various degeneracies can be present or suppressed depending on model priors used in the reconstruction.}

We already noted that the difference in Group B vs. C is entirely due to the inclusion of additional data in the latter; the modeling assumptions are the same for both. SPEMD differ from other models most likely because they are constrained by the model priors to be power laws, thereby suppressing degeneracies that would result in other profile shapes.

The most interesting comparison is between \glass~and Group B models, both of which have nearly identical data inputs. Despite having quite different modeling approaches, pixelated free-form vs. two component parametric model, their density profiles interior to the image radius are very similar in all 4 panels. This is encouraging because {it means that if the same data is used in combination with flexible mass models then the radial lensing degeneracies {can be} largely broken.} 

For the models that differ, it is interesting to consider what type of lensing transformation connects their density profiles. The shapes of the profiles indicate that the transformation that would map any one of these profiles into another is not the mass sheet degeneracy (MSD). For example, applying a MSD transformation to the power law \hc~model, one obtains a profile with lower central normalization, and profile shape which is convex, especially at larger radii. But the \glass~and our models, whose central normalization is lower than that of \hc, have concave outer profile shapes. Therefore transformations mapping these model profiles onto each other belong to a different family of source position transformations \cite[][SPT]{sch14}, not MSD. This means that MSD transformation is not the only radial degeneracy one should consider when modeling lenses.

\subsubsection{Density profile within $r\sim 100\,$pc}

In \wfi, the size of the central density core is not constrained by the four images, leading to very different central densities of the models in Figure~\ref{fig:Comparison} and \ref{fig:mixed}. If the central fifth image were detected in \wfi, it would provide a useful constraint at the lens center. Unfortunately, central images are rarely detected. There are only two known detections in lenses with a single lensing galaxy. \cite{win2003} use radio observations to detect the central image in PMN J1632-0033, which is demagnified to $0.004$ of the brightness of the two bright images. \cite{mul2020} use ALMA observations to detect the central image, demagnified to $0.007$ of the brightness of the other two images in the double lens, PKS 1830-211. Upper limits have also been measured. For example, \cite{qui2016} place an upper limit of $\sim 10^{-4}$ on the magnification of the central image with respect to the brightest visible image in a double image system B1030+074. 

Our models, as well as those from \glass, do not have that level of demagnification. For our Group C models, it is typically 0.013, while the smallest is $1.8\times 10^{-3}$. However, all models presented here can be made compatible with a highly demagnified central image by reshaping the central mass distribution in a monopole-like fashion \citep{sah00,lie12}, without altering model predictions for the four observed images. 
Starting with the average profile of Group C, and then shaving off 0.13\% (0.6\%) of the surface density from the annulus interior to the images (0.1"--1"), and using it to form a flat density core in the inner 100 pc ($0.0144"$) will demagnify the central image to $10^{-3}$ ($10^{-4}$) compared to the brightest, second arriving image. 

To demagnify to the level of $10^{-3}$ would make the central slope outside of the 100~pc core somewhat steeper, but still shallower than in the outer regions, preserving the concave nature of the profile. Adding a supermassive back hole (SMBH), which must be present, would further demagnify the image, often rendering it invisible \citep{qui2016}.

\subsection{Azimuthal mass distribution and lopsidedness}


Given the results of \cite{rusu2020h0licow}, it appears to be impossible to fit the quad point images with an elliptical galaxy mass distribution, whether it consists of one or two co-centered mass components, even if the mass center is allowed to be displaced from the light center, and G2 is allowed to have flexible ellipticity and position angle. Some additional modeling freedom in the azimuthal direction appears to be required. For example, the \glass~models, which reproduce point image properties exactly, deviate significantly from azimuthal elliptical symmetry.

Our models allow extra azimuthal freedom in the form of offset centers. However, unlike the \glass~models, the deviation from ellipticity in all three sets of our models are small (Figure~\ref{fig:AverageConvergence}).
Without the contribution from G2, fractional deviations from inversion symmetry inside the Einstein radius are of order few percent. Apparently, in the case of \wfi~even such small deviations are sufficient for the purpose of forcing the mass and light centroids of the main lens to coincide, and the image positions, time delay ratios and flux ratios to be fit within observational errors.  

Since the degree of lopsidedness is small, it is possible that LoS can be partly responsible, as it is not expected to be distributed elliptically symmetrically with respect to the lens center. In fact, LoS structure can result in lopsided mass distribution, with fractional mass excess of $\lesssim 1\%$ to a few percent \citep{veg14,gil19,gil20}. However, the fraction of lensing lines of sight that contain massive LoS substructure ($\gtrsim 10^9M_\odot$) is about 10\% \citep{gra18,veg14}. Unfortunately, there is no robust way to know if a small fraction of lensing mass is part of the LoS or the main lens galaxy. The former would result in non-diagonal magnification matrix, but applying that test to observations would be difficult \citep{cag20}.

While we do not know the origin of lopsidedness in \wfi, for lenses with low lens and source redshifts, like \iptf, with $z_l=0.22$ and $z_s=0.41$ \citep{goo17}, LoS is unlikely to be responsible, and the lopsidedness found in \cite{williams2020two} is most likely intrinsic to the lens plane.

\subsection{Lensing degeneracies involving external shear}

A qualitative comparison between the three models in Figure~\ref{fig:Comparison} reveals that the orientations of the mass distributions are approximately the same, elongated towards the last arriving saddle image. Small differences in the position angle can be attributed to the degeneracy between ellipticity and external shear. This well known degeneracy is easy to understand because both ellipticity and shear result in similar effects on the image positions around the lens center. 

\begin{figure}
    \centering
    \vspace{0.75cm}
    \includegraphics{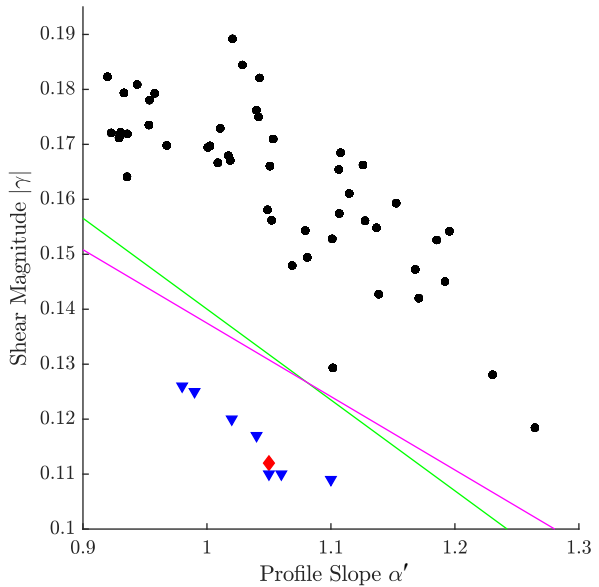}
    \caption{The magnitude of external shear vs. the mean lensing potential slope {$\alpha^\prime$ (not the same as $\alpha$ in eq.~\ref{eq:alphapot})} of the main lensing galaxy in the annulus {centered on the average image radius, 1.16", and of width $\pm 25\%$}. Recall that {$\rho_{2D}\propto r^{\alpha^\prime-2}$}. The black points represent 50 models of our Group C. The red rhombus is the main SPEMD model from \cite{rusu2020h0licow}, while the blue triangles are the other SPEMD models from their Table~C1. \glass~model is off the limits of the plot at $|\gamma|=0.225$ and {${\alpha^\prime}=0.37$}. Best-fit lines for our models from Group A and B are shown in green and magenta, respectively. There is considerable scatter around the lines, so we do not show the individual points to avoid overcrowding. Overall, shallower density profiles require less external shear.}
    \label{fig:shearslope}
\end{figure}

The lesser appreciated degeneracy is between the magnitude of external shear and the slope of the lens galaxy density profile. Figure~\ref{fig:shearslope} shows that for our Group C models (black points) shallower density profiles require less external shear to reproduce the images. {The potential slope $\alpha^\prime$ in that Figure corresponds to the superposition of the two mass components.} The same relation applies to our Group A and B models (green and magenta best-fit lines, respectively), and to the SPEMD models (red and blue points), but with a different normalization. The reasons for the difference in amplitude of the relation is not clear; it could be due to a combination of different modeling priors and different data sets. Here, we will concentrate on the slope of the relation.

The anti-correlation between $\alpha$ and $\gamma$ can be understood using a simplified model consisting of a circular power law lensing potential with external shear, 
\begin{equation}
    \Psi=b r^\alpha+\frac{1}{2}\gamma r^2\,\cos2(\theta-\theta_0), 
\end{equation}
with $\theta_0=0$, and source which is on the same axis as the shear, at distance $x_\beta$ from the center of the lens. Let the two images form at $x_1$ and $x_2$. The lens equation, $x-x_\beta=\partial\Psi/\partial x$ can be written for the two images as, 
$x_1-x_\beta= \alpha b x_1^{\alpha-1}+\gamma x_1$ and
$x_2-x_\beta=-\alpha b x_2^{\alpha-1}-\gamma x_2$.
The radial separation between the two images in this simplified case is
\begin{equation}
    \Delta|x|=|x_1|-|x_2|=2b(x_\beta/b+\alpha\gamma)/(1-\gamma^2),
\end{equation}
where we have assumed that the dependence on $\alpha$ in the exponent in the lens equation can be neglected because $\alpha$ is close to 1. One of the main effects of shear is to increase this separation between images. Assuming that the separation in doubles and quads behaves similarly, we can obtain the dependence on $\alpha$ and $\gamma$ from $\Delta|x|$. Since $x_\beta$ is always small compared to the Einstein radius, $b$, and $(1-\gamma^2)\approx 1$, we obtain that $\Delta|x|\propto \alpha\gamma$. (Recall that projected mass density scales as $\rho_{2D}\propto r^{\alpha-2}$.) Because all viable models have to maintain the same radial separation between images, we expect that the radial slope of the potential, $\alpha$, will be anti-correlated with the magnitude of external shear, $\gamma$. This is what is seen in Figure~\ref{fig:shearslope}, {where each set of solutions using the same data and the same priors are described by this anti-correlation.}



\vspace{0.5cm}

\subsection{Comparison in model space}\label{sec:compmod}

In the previous subsections we compared specific physical properties of galaxy models reconstructed by different techniques. Here we ask a somewhat different question, aimed at techniques capable of exploring large regions of model space by virtue of having more parameters than observables: is there any overlap between solutions of methods that have very different assumptions but use the same data? 

Traditional parametric methods, like those in \cite{rusu2020h0licow}, are not useful for this purpose, because their models have more observables than parameters, and hence obtain only one solution, with statistical uncertainties. This analysis is confined to our models and \glass. (An additional reason to restrict our comparison to these two methods is that these reproduce all the properties of the quasar point images to observational precision.)

For convenience, we do the comparison only at the locations corresponding to \glass~pixel centers.  We discretized our mass distributions into a grid with the same number of data locations as the \glass~grid. This amounts to about $n\sim 150$ grid locations over the whole lens plane of radius $1.478''$, which contains the largest image radius, $1.472''$.

\begin{figure}
    \center
    \vspace{0.05cm}
    \includegraphics[scale=1]{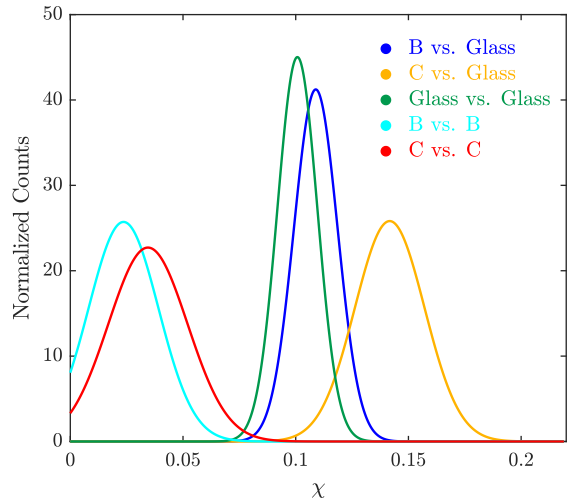}
    \caption{Distributions resulting from computing $\chi$ between pairings from different model populations; see Section~\ref{sec:compmod}. Because the distributions resulting from each of these five tests were very close to Gaussians, we just plot the fitted Gaussians. The distributions from models A are omitted because they are very similar to those from models B.}
    \label{fig:hist}
\end{figure}

To compare any two models, we arrange their $\kappa$ values into two $n$-dimensional vectors $\vec{x}$ and $\vec{y}$, and evaluate their ``similarness" by computing  $\chi=|\vec{x}-\vec{y}|/|\vec{x}+\vec{y}|$, which compares the lengths of the difference and sum of vectors $\vec{x}$ and $\vec{y}$. This was done for each of the $50\times1000$ pairs of our models B, and C vs. \glass, the same number of random pairings of \glass~vs. \glass~models, and $50\times 50$ pairs of B vs. B, and C vs. C models. The distributions resulting from each of these five tests were very close to Gaussians (probably because of the central limit theorem), so in Fig.~\ref{fig:hist} we just plot the five fitted Gaussians. (Results for A models are not shown because they are very similar to those of B.) 

Let us first consider B vs. B distribution, represented by the cyan curve. The distribution has some overlap with $0$, implying that most B models are quite similar to each other. A pair of identical models has $\chi=0$.
The most dissimilar ones differ from each other by a fractional difference of only $\sim 5-7\%$, therefore in $n-$dimensional space these form a tight cloud of models. Models C (red) show similar traits, but these form a somewhat larger cloud. \glass~(green) forms a more dispersed cloud, with models typically separated by a fractional difference of $\sim11\%$. The fact that the mean of the \glass~distribution is larger that that of the B distribution is not surprizing, since the former a free-form method, while the latter is parametric.  The narrowness of \glass~distribution means that the models are approximately equidistant from each other.  

The main question posed in this subsection is answered by the blue and gold curves. B and \glass~models (blue) are typically separated by the same distance as two models from the \glass~ensemble. Combined with the tightness of the B vs.~B model cloud, this suggests that B models occupy a region of model space close to the \glass~cloud, and that there is likely a small overlap between the two. However, a more in-depth analysis, for example using Principle Components (PCA) would need to be carried out for a more thorough understanding of the relative distributions of the models in these high-dimensional spaces. This is consistent with our earlier finding that B and \glass~models have a very similar radial density profile (see Section~\ref{sec:denprof} and Figure~\ref{fig:mixed}). 

We conclude that methods using very different parametrizations of the lensing mass---like ours vs. \glass---if they allow for degenerate mass models and use the same data, can converge on similar solutions, i.e. their solutions overlap in model space.

If the data is different, as is the case with our models C vs. \glass~(gold curve), then the resulting solutions are different, with little or no overlap.


\subsection{Assigning lensing mass to astrophysical entities: interpretation degeneracy}\label{sec:interp}

Lensing responds to all the mass between the source and the observer, whether it belongs to the main lensing galaxy, other nearby galaxies, or LoS structures. Free-form modeling, like \glass, takes advantage of that, and is completely agnostic to the physical origin of the deflecting mass. The downside is that it makes it harder to draw conclusions about the mass distribution in the main galaxy. For example, \glass~model in the left panel of Figure~\ref{fig:Comparison} shows two main deviations from ellipticity: the extension of density contours towards the lower left, which is most apparent around the image radius, and the extension towards the upper left, within 1-2 pixels from the center. The former must be, at least partially due to the mass of G2, which is located in that direction. The latter could be due to the asymmetry in the main lensing galaxy, or due to G2, or both. The contribution from the LoS structures is also probably present at some level.

While FF modeling provides the clearest example of our inability to uniquely and correctly interpret how to assign mass to various astrophysical entities, this `interpretation degeneracy' is not limited to FF modeling. In the SP approach, even if the modeler intended that a certain fraction of the total projected mass belongs, say, to the secondary galaxy, it maybe well be part of the main galaxy, or vise versa. For example, in SPEMD models the orientation of the ellipticity of G2 is such that it contributes more mass to the lower portion of the quad (in the orientation shown in Figure~\ref{fig:Comparison}), around the second arriving image, than a circular G2 would. It is possible, however, that this mass asymmetry due to the model ellipticity of G2 should be attributed to the main lens, or even LoS mass.

In our own models, as we already mentioned, the physical origin of the offset mass component, or part thereof, is unclear. It could be intrinsic to the main lens, or, since \wfi's lens and source are high,  some of it could arise from LoS structures.

While the most problematic consequence of mis-assigning mass to the wrong entity is for the physical understanding of the lensing galaxy, other aspects of lensing related analysis may be affected as well.  When single-aperture stellar kinematics are used, interpretation degeneracy can make the analysis less self-consistent, because the Jeans equation combines the radial mass distribution, some of which may be incorrectly attributed to the main lensing galaxy, with the velocity dispersion, which is measured from the main galaxy only.  It is possible that some of these issues will be alleviated with the use of spatially resolved stellar kinematics \citep{yil20}.



\section{Conclusions}\label{sec:conc}

Most recent papers modeling galaxy-scale strong lensing are focused on estimating the Hubble parameter, $H_0$, because of unresolved and tantalizing discrepancy between the values derived from various independent probes \citep[e.g.,][]{verde2019tensions,riess2020expansion}. This puzzle will ultimately be resolved, and the goal of galaxy-scale lens modeling will shift back to the detailed structure and evolution of the lensing galaxies. 

Our motivation for choosing \wfi~is that it is one of six \hc~systems, and its quasar point images and lens light center are poorly fit by their power law and composite models. This can be interpreted as evidence that the combination of quasar image data, extended ring data and galaxy velocity dispersion are not compatible with the power law, or composite profiles. 
{In the interest of working around some of the shortfalls of this modelling strategy, we introduce more flexible models, which offer more freedom while still being built out of astrophysically motivated components. We do not include extended ring image and stellar velocity dispersion as constraints.}

Our goal is to determine what can currently be said about the structure of the lens galaxy in \wfi. Aware of the fact that conclusions about mass distribution can depend on the lens modeling assumptions, we examine models derived from three independent modeling techniques, simply parametrized (SP), pixelated free-form (FF), and our own models that bridge the gap between the SP and FF modeling philosophies. Our use of two superimposed simply parametrized lensing potentials is similar to the SP approach, but the freedom afforded by our use of offset centers, and, as importantly, by the fact that we generate many models, instead of one, is reminiscent of the FF approach. (We note that our way of parametrizing the lens mass is not unique, and other parametrizations could give rise to somewhat different models.)

We present 3 groups of models, sequentially including {constraints due to quasar image positions} (Group A), time delay ratios (Group B), and flux ratios (Group C), assuming they are due to macro model  (their microlensing origin is already ruled out). We find that time delay ratios in \wfi~do not constrain models any more than image positions already do, while flux ratios make the central density slope significantly steeper, but still shallower than isothermal.

Despite the emphasis placed in the literature on the mass sheet degeneracy and the corresponding transformation, we show that the transformations connecting SP, FF and our models are not MSD, but more generalized source position transformations, SPT \citep{sch14}. In the absence of lensing constraints in the central $\sim 100$~pc, the current models have a wide range of central densities. This is not a real disagreement because the mass in the central regions can be reshaped using monopole transformations, without affecting the fit to the four images of the quad.  In the future, the detection of, or an upper limit on the flux of the central odd image will help constrain the central density and hence the density profile shape. 

In addition to the well known ellipticity-shear degeneracy, the models we examine manifest a degeneracy between the magnitude of external shear and the lens profile slope. Each set of models, for example, SPEMD, or our Group C, show an anti-correlation between these quantities, but the normalization of the anti-correlation differs between SPEMD, Group C, etc., {for reasons that are not yet clear but could depend on the data sets used, or model priors.}

Having examined various aspects of the three different sets of models of \wfi---power laws, pixelated free-form, and our own flexible parametrized models---we conclude that there are as many similarities as there are differences between them. The similarities are mostly in the azimuthal structure of the lensing mass: though details differ, all models agree that the mass distribution is elliptical, their position angles are within $\sim30^\circ$ of each other, and that there has to be significant external shear, $\gamma\gtrsim 0.1$. There also has to be mass extension beyond ellipticity in the direction between the 2nd and 4th arriving images. There is less agreement between models on the density profile slope, both at the center and around the image annulus. {The only exception is that of \glass~and our Group B models, whose radial profiles look similar, possibly because they share the same data constraints and considerable modeling freedom.}  Future observations, like spatially well resolved line of sight velocity dispersion of the lens galaxy, and limits on the flux of the central odd image will help better constrain the lensing mass distribution and determine its astrophysical origin.

The three sets of our models, Groups A, B and C require the lens mass distribution to be lopsided. {The excess mass arising from the lopsidedness (compared to inversion symmetric, or, SP models) is only a few percent. Therefore we find that in \wfi~}it does not take much lopsidedness to generate $\chi^2\leq1$ models, i.e., to make predicted image positions, time delay ratios and flux ratios to be consistent with observational uncertainties, and to make galaxy light center agree with the mass center. The astrophysical origin of lopsidedness in the case of \wfi~ is not clear: it could arise in the main lensing galaxy, or be contributed by nearby galaxies, or the line of sight structures. This inability to correctly assign mass to its astrophysical origin may not affect the estimation of $H_0$, but it does affect the interpretation of galaxy structure. For example, if the lopsidedness does arise in the main lens then this $z=0.66$ galaxy is not well relaxed in the innermost few percent of its virial radius. 

This is an example of what we call interpretation degeneracy. As opposed to the lensing degeneracies, which allow different mass models to reproduce the same lensing observables, the interpretation degeneracy, which stems from the fact that lensing responds to all mass, applies to any given mass model. It makes it difficult to determine how much of the model's mass is due to the main lensing galaxy vs. other nearby galaxies vs. line of sight structure. Interpretation degeneracy is not a new notion in lens modeling \citep[e.g.,][]{tro00}, and will likely become more important when the main goal of modeling becomes the detailed structure of intermediate redshift galaxies.

\begin{acknowledgments}
The authors would like to thank Prasenjit Saha for suggesting tests that extended the paper's analysis into new directions and the REU program at the University of Minnesota, supported by grant NSF-1757388. 
\end{acknowledgments}

\appendix

\section{Initial Parameter Ranges}\label{app:Clusters} 

\begin{figure}[h!]
    \centering
    \includegraphics{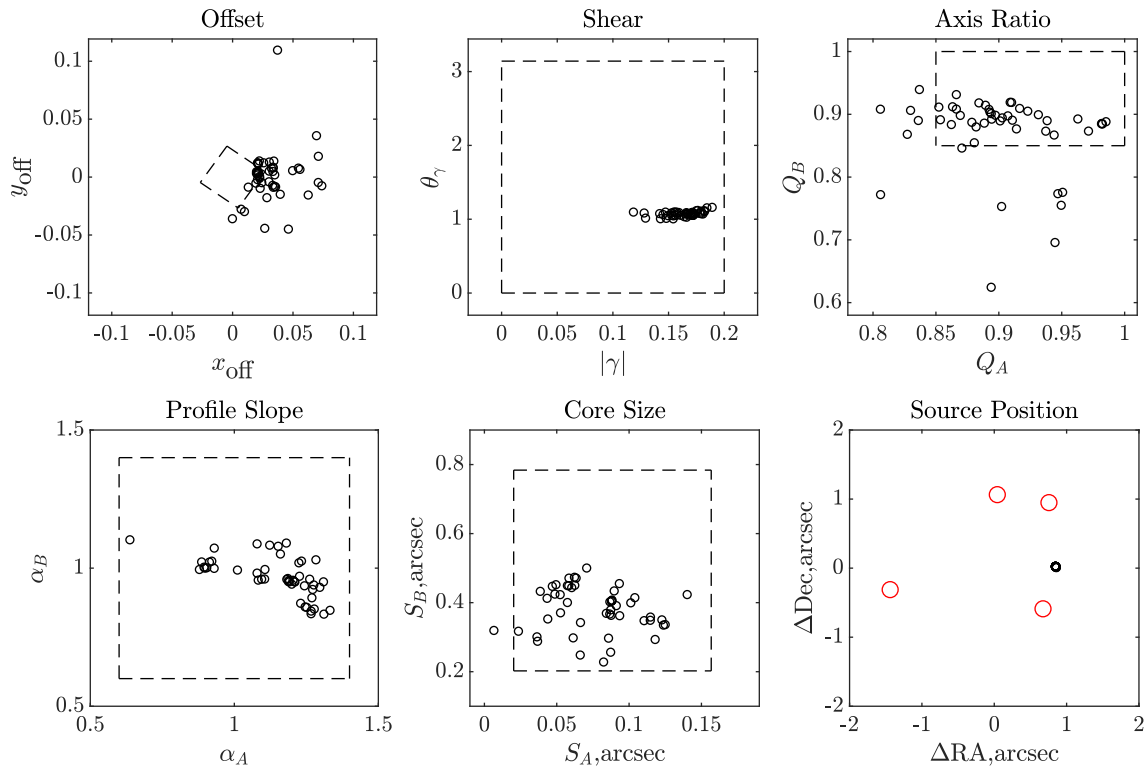}
    \caption{Final parameters for the 50 models of Group C; see eq.~\ref{eq:alphapot}-\ref{eq:QQQ}. The dashed boxes show the limits within which initial parameters were randomly scattered for all three Groups. The dashed box in the upper left panel is rotated because the Cartesian coordinate system where calculations were carried out was rotated with respect to the orientation of figures in this paper.}
    \label{fig:Clusters}
    \vspace{0.35cm}
\end{figure}

Figure \ref{fig:Clusters} shows parameter ranges for 50 models from Group C. The dashed black boxes correspond to the initialization ranges, which were the same for all three Groups A, B and C. Even though these ranges are somewhat arbitrary, the dependence on them is partly erased in the final models because parameter optimization using downhill simplex can wander outside of the initial ranges. In fact, in the case of the offset between the center of the two components (upper left panel), the bulk of the final models are outside of the initial box. That box was made purposefully small around $(0,\,0)$ to force the two mass components of the model to have coincident centers as much as possible. However, models appear to prefer $r_{\rm off}\gtrsim 0.025"$, or $0.174$~kpc. Source positions are also shown in the last panel, along quasar image positions, depicted by red circles. 

\bibliography{WFI2033}
\end{document}